\documentclass[useAMS,usenatbib]{mn2e} 

\usepackage{graphicx}
\usepackage{ulem}

\newcommand{\kms}{km s$^{-1}$}
\newcommand{\degree}{$^{\circ}$}

\title[Nova Oph 2009]{Properties, evolution and morpho-kinematical modelling
of the very fast nova V2672 Oph (Nova Oph 2009), a clone of U Sco}
\author[U. Munari et al.]{U. Munari$^{1,2}$, V. A. R. M.
Ribeiro$^{3}$, M. F. Bode$^{3}$ and T. Saguner$^{1,4}$\\
$^{1}$INAF Astronomical Observatory of Padova, via dell'Osservatorio, 36012
Asiago (VI), Italy\\
$^{2}$ANS Collaboration, c/o Astronomical Observatory, 36012 Asiago (VI),
Italy\\
$^{3}$Astrophysics Research Institute, Liverpool John Moores University,
Twelve Quays House, Egerton Wharf, Wirral, \\ Birkenhead, CH41 1LD, UK\\
$^{4}$Department of Astronomy, University of Padova, Vicolo
dell'Osservatorio 5, 35100 Padova, Italy
}
\begin{document}

\date{Accepted 1988 December 15. Received 1988 December 14; in original form
1988 October 11}

\pagerange{\pageref{firstpage}--\pageref{lastpage}} \pubyear{2002}

\maketitle

\label{firstpage}

\begin{abstract}
Nova Oph 2009 (V2672 Oph) reached maximum brightness $V$=11.35 on 2009
August 16.5.  With observed $t_2$(V)=2.3 and $t_3$(V)=4.2 days decline
times, it is one of the fastest known novae, being rivalled only by V1500
Cyg (1975) and V838 Her (1991) among classical novae, and U~Sco among the
recurrent ones.  The line of sight to the nova passes within a few degrees
of the Galactic centre.  The reddening of V2672 Oph is $E_{B-V}$=1.6
$\pm$0.1, and its distance $d$$\sim$19 kpc places it on the other side of
the Galactic centre at a galacto-centric distance larger than the solar one. 
The lack of an infrared counterpart for the progenitor excludes the donor
star from being a cool giant like in RS~Oph or T~CrB. With close
similarity to U~Sco, V2672 Oph displayed a photometric plateau phase, a He/N
spectrum classification, extreme expansion velocities and triple peaked
emission line profiles during advanced decline.  The full width at zero
intensity of H$\alpha$ was 12,000 \kms\ at maximum, and declined linearly in
time with a slope very similar to that observed in U~Sco.  The properties
displayed by V2672 Oph lead us to infer a mass of its white dwarf
close to the Chandrasekhar limit and a possible final fate as a SNIa. 
Morpho-kinematical modelling of the evolution of the H$\alpha$ profile
suggests that the overall structure of the ejecta is that of a prolate
system with polar blobs and an equatorial ring.  The density in the prolate
system appeared to decline faster than that in the other components. 
V2672 Oph is seen pole-on, with an inclination of 0$\pm$6$^\circ$ and an
expansion velocity of the polar blobs of 4800$^{+900}_{-800}$~\kms.  On the basis of
its remarkable similarity to U Sco, we suspect this nova may be a recurrent. 
Given the southern declination, the faintness at maximum, the extremely
rapid decline and its close proximity to the Ecliptic, it is quite possible
that previous outbursts of V2672 Oph have been missed.  
\end{abstract}


\section{Introduction}

Nova Oph 2009 (= V2672 Oph) was discovered by K. Itagaki on August 16.515 UT at
mag 10.0 on unfiltered CCD images \citep[cf.][]{nakano}, at
position (J2000) $\alpha$ = 17$^h$38$^m$19$^s$.70, $\delta$ =
$-$26$^\circ$44$^\prime$14$^{\prime\prime}$.0. This corresponds to Galactic
coordinates $l$=001.021, $b$=+02.535, thus V2672 Oph lies very close 
to the direction to the Galactic centre.

Spectroscopic confirmation was obtained on August 17.6 UT by \citet{AMH09}
and on August 17.8 UT by \citet{MSO09}.  Both reported a  very
large velocity width for H$\alpha$ and the paucity of other emission lines
on a red and featureless continuum.  \citet{MSO09} noted the great strength
of interstellar lines and diffuse interstellar bands, which indicated a
large reddening and therefore accounted for the red slope of the continuum
of the nova.  \citet{MSO09} also noted how the profile of the emission lines
was highly structured, in the form of a broad trapezium with a sharper
Gaussian on top.  The trapezium had a velocity width of $\sim$11700 \kms\ at
the base and $\sim$6900 \kms\ at the top, while the sharp Gaussian component
had a FWHM of $\sim$1400 \kms.  They also noted how the colours, the rapid
decline, and the velocity of the ejecta suggested that V2672 Oph was an
outburst occurring on a massive white dwarf, not dissimilar to the U~Sco
type of recurrent novae, and that a search in plate archives for missed
previous outbursts could pay dividends.

A report on an early X-ray detection and following evolution of V2672 Oph
was provided by \citet{SOP09}.  Their August 17.948 observation
with the {\it Swift} satellite detected the nova with both the XRT and
UVOT instruments.  Further observations on the following days found the nova
to  emit at a stable X-ray flux level, while rapidly declining at
ultraviolet wavelengths.  \citet{SOP09} suggested that the early hard X-ray
emission was likely due to shocks between the fast ejecta and a pre-existing
circumstellar medium (as in the recurrent nova RS Oph $-$ e.g. 
\citealt{BOO06}) or intra-ejecta shocks \citep[as in the very fast
classical nova V838 Her,][]{OLB94}.  \citet{SOP09} also looked for past
X-ray observations in the field, and concluded that V2672 Oph was not
recorded as an X-ray source prior to the 2009 outburst.  They also noted how
the nova was not detected in gamma-rays by INTEGRAL/IBIS during Galactic
bulge monitoring observations taken on 2009 August 20 and 23/24.

\begin{table}
\centering
\caption{Our $B$$V$$R_{\rm C}$$I_{\rm C}$ photometry of V2672 Oph. The
first column is the UT date in 2009 August. The second column gives the 
heliocentric Julian day $-$ 2455000.}
\begin{tabular}{@{}ccccc@{~~}c@{~~}c@{}}
\hline
date & HJD & $V$ & $B$$-$$V$ & $V$$-$$R_{\rm C}$ & $V$$-$$I_{\rm C}$ 
     & $R_{\rm C}$$-$$I_{\rm C}$\\
\hline
         17.473 & 60.9727 & 12.466  & 1.777  & 1.619  & 2.737  & 1.137  \\
         17.801 & 61.3014 & 12.722  & 1.772  & 1.596  & 2.699  & 1.075  \\
         18.499 & 61.9988 & 13.200  & 1.735  & 1.677  & 2.619  & 0.977  \\
         18.816 & 62.3162 & 13.320  & 1.770  & 1.694  & 2.542  & 0.906  \\
         20.820 & 64.3196 & 14.417  &        & 1.652  & 2.445  & 0.864  \\
         22.834 & 66.3336 & 15.871  &        &        & 2.684  &        \\
         23.819 & 67.3193 & 15.839  &        & 1.600  & 2.417  & 0.849  \\
         24.823 & 68.3231 & 15.923  &        & 1.421  & 2.215  & 0.819  \\
         25.816 & 69.3160 & 16.342  &        & 1.529  & 2.351  & 0.846  \\
\hline
\end{tabular}
\label{tb:tab1}
\end{table}

\citet{HRM09} reported their detection of radio emission from V2672 Oph
during the first two weeks after optical maximum.  The radio emission from
most novae is dominated by thermal bremsstrahlung which is optically thick
at early times, leading to a dependence of the flux on frequency as $F_\nu
\propto \nu^{+\alpha}$, where $\alpha \sim$ 1 $-$ 2 at early times. 
They detected V2672 Oph at 8.46 GHz with the VLA on Sept 1.13, but obtained
no detection at 22.46 GHz two days later, on September 3.18.  \citet{HRM09}
concluded that this was best explained by a synchrotron origin for the radio
emission observed from V2672 Oph.  To support this view, they noted that
($i$\,) the strong shocks in its ejecta, suggested by the hard X-ray
emission, can also be the source for the relativistic electrons and strong
magnetic fields needed to generate synchrotron radiation, and that ($ii$\,)
the recurrent nova RS Oph, shows strong radio synchrotron emission within
days of the outburst \citep[e.g.,][]{PDB85,OBP06,EOB09}.

The peculiarity and rarity of the phenomena displayed by V2672 Oph is
evident from these early preliminary accounts.  In this paper we report our
optical observations, that have allowed us to derive the photometric and
spectroscopic evolution and the basic properties of V2672 Oph, and
which are used to perform morpho-kinematical modelling of the emission line
profiles and hence disentangle the basic components of the expanding
ejecta.

\section{Observations}

Photometric observations of V2672 Oph have been obtained with two
instruments in collaboration with S.  Dallaporta, A.  Frigo, A.  Siviero, S. 
Tomaselli, A.  Maitan and S.  Moretti of ANS Collaboration.  The first is a
0.25-m Meade LX-200 Schmidt-Cassegrain telescope located in Cembra (Trento,
Italy).  It is equipped with an SBIG ST-8 CCD camera, 1530$\times$1020
array, 9~$\mu$m pixels $\equiv$ 0.74$^{\prime\prime}$/pix, with a field of
view of 19$^\prime$$\times$13$^\prime$.  The $B$$V$$R_{\rm C}$$I_{\rm C}$
filters are from Schuler.  The other one is a 0.25-m f/6 Ritchey-Chretien
robotic telescope, part of the GRAS network (GRAS15, Australia).  It carries
an SBIG ST-10XME CCD camera 2184$\times$1472 array, 6.8 $\mu$m pixels
$\equiv$ 0.93$^{\prime\prime}$/pix, with a field of view of
34$^\prime$$\times$23$^\prime$.  The $B$$V$$R_{\rm C}$$I_{\rm C}$ filters
are again from Schuler.  The calibration of photometric zero points and
colour equations have been carried out against the \citet{L83,L92,L09}
equatorial standards.  The photometric data are presented in Table~1.  The
total error budget (dominated by the Poissonian noise, with only a minor
contribution from the uncertainty of the transformation to the Landolt
system of equatorial standards) does not exceed 0.035 mag for all points.

Spectroscopic observations of V2672 Oph were obtained with two
telescopes, in collaboration with P.  Valisa, V.  Luppi and P.  Ochner of
ANS Collaboration.  A journal of the observations is given in Table~2.  The
0.6m telescope of the Schiaparelli observatory in Varese (Italy), is
equipped with a multi-mode spectrograph (Echelle + single dispersion modes)
and a SBIG ST10-XME CCD camera (2184$\times$1472 array, 6.8~$\mu$m pixels). 
It was used to obtain low resolution, wide wavelength range spectra.  The
Asiago 1.22m telescope feeds light to a B\&C spectrograph, equipped with a
1200 ln/mm grating and ANDOR iDus 440A CCD camera (EEV 42-10BU back
illuminated chip, 2048$\times$512 pixels, 13.5~$\mu$m in size).  It was used
to obtain the higher resolution spectra around H$\alpha$.  The spectra
collected with both telescopes were calibrated into absolute fluxes by
observations of several spectrophotometric standards, which were observed at
air-masses close to those of V2672 Oph.  The zero points and slopes of
the absolutely fluxed spectra were checked against the photometry of
Table~1, by integrating the spectral flux through the $B$$V$$R_{\rm C}$
photometric bands.  The error on the fluxes of our spectra turned out not to
exceed 9\% over the wavelength range covered.

 \begin{table}
 \centering
 \caption{Journal of the spectroscopic observations. The time in the third
          column is counted from maximum brightness on August 16.515 UT.}
          \begin{tabular}{@{~}rccc@{~~}c@{~~}cc@{~}}
             \hline
 \multicolumn{1}{c}{date} &  UT & $\Delta t$ & expt & disp & $\lambda$ range 
              & tel.  \\
              &       & (days) & (sec) &    (\AA/pix)& (\AA)           \\
             \hline
     Aug   17 & 19:52 & +1.31 & 1800  & 0.60         & 5575-6805 & 1.22m\\
           17 & 20:28 & +1.34 & 1500  & 2.12         & 4000-8625 & 0.6m \\
           18 & 20:31 & +2.34 & 1200  & 0.60         & 5575-6805 & 1.22m\\
           19 & 20:28 & +3.34 & 2400  & 0.60         & 5590-6810 & 1.22m\\
           19 & 20:56 & +3.36 & 1800  & 2.12         & 4010-8650 & 0.6m \\
           20 & 20:11 & +4.33 & 1800  & 4.24         & 4010-8650 & 0.6m \\
           21 & 19:52 & +5.31 & 1800  & 0.60         & 5540-6775 & 1.22m\\
           24 & 20:11 & +8.33 & 3600  & 0.60         & 5525-6760 & 1.22m\\
             \hline
 \end{tabular}
 \label{tb:tab2}
 \end{table}

\section{Photometric evolution}

The photometric evolution of V2672 Oph is presented in Figure~1.
The light- and colour-curves of Figure~1 were obtained by combining 
our data in Table~1 with other sources as follows. 

The unfiltered CCD photometry obtained around the time of discovery by
various Japanese observers and reported by \citet{nakano}, was scaled
to $V$ band by adding +1.35 magnitudes.  The Japanese amateurs usually
calibrate their unfiltered photometry against the $R_{\rm C}$ values of
field stars taken from the USNO-B catalog.  As illustrated in Table~1
and Figure~1, $V$$-$$R_{\rm C}$ displayed by V2672 Oph has remained
fairly stable over the initial photometric evolution.  Thus the application
of the same shift to all data by \citet{nakano} seems reasonable.  The
amount of the shift has been derived by comparison with simultaneous $V$
band values from Table~1.

Members of the VSNET\footnote{http://www.kusastro.kyoto-u.ac.jp/vsnet/}
Japanese amateur organisation performed some CCD filter photometry.  This
photometry is usually not corrected for colour equations and linked to
non-specified calibration stars.  Given the very red colours of V2672 Oph,
this introduces large shifts.  By comparison with the photometry in Table~1,
we found the following correction to VSNET data: $B$ = $B_{\rm VSNET}$ +
0.31, $V$ = $V_{\rm VSNET}$ + 0.29, $R_{\rm C}$ = $R_{\rm VSNET}$ + 0.10,
$I_{\rm C}$ = $I_{\rm VSNET}$ + 0.05.  We have applied these to the VSNET
data and plotted them in Figure~1.

 \begin{figure}
 \includegraphics[width=85mm]{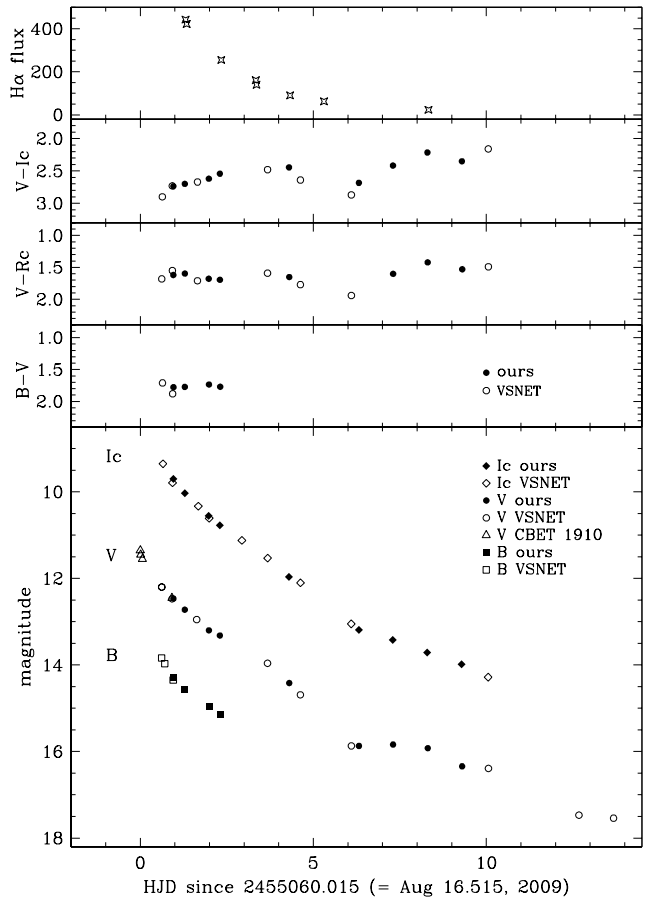}
 \caption{Photometric evolution of V2672 Oph. The integrated 
 H$\alpha$ flux in the top panel is in units of 10$^{-13}$ erg cm$^{-2}$ 
 s$^{-1}$ \AA$^{-1}$.}
 \label{fig:fig1}
 \end{figure}

\subsection{Maximum and early decline}

V2672 Oph was probably discovered quite close to maximum brightness. The
earliest observations given by \citet{nakano} report the nova at
$V$=11.35, 11.45 and 11.55 on August 16.515, 16.526 and 16.576,
respectively.  This would correspond to an extremely fast decline of 3.3 mag in 1
day, something that U~Sco (the fastest known recurrent nova, \citealt{S10})
displays only during the first hours past true maximum, before slowing down
to a decline time $t_{2}(V)$$\approx$2.0 days \citep{MDC10}.  A patrol
observation on August 14.142 (2.373 days before discovery,
\citealt{nakano}) recorded the nova fainter than $V$=14.0 limiting
magnitude.  In the rest of this paper, we will assume that the time of
discovery, August 16.515, coincides with the time of maximum optical
brightness.

The lightcurve of V2672 Oph in Figure~1 is characterized by a smooth
behaviour and decline times
\begin{equation}
t_2(V)=2.3   ~~~~~~~ t_3(V)=4.2  ~{\rm days}
\end{equation}
($\pm$0.1 days) which are close to $t_2(V)$=2.0, $t_3(V)$=4.2 that are the
mean values for the 1999 and 2010 outbursts of U~Sco \citep{MZT99,MDC10}. 
V2672 Oph therefore qualifies as one of the fastest known novae. 
Among classical novae, similar or greater speeds have been attained only by
V1500 Cyg (1975) and V838 Her (1991), while among recurrent novae, only
U~Sco could rival V2672 Oph.

After  declining past $t_3(V)$, V2672 Oph displayed an inflection in its
light-curve, easily visible in Figure~1, and far more pronounced in $V$ than
in $I_{\rm C}$.  Around day +4.5, the nova started to decline faster than
expected from smooth extrapolation of the preceding portion of the
light-curve.  The maximum deviation was reached around day +5.8, when V2672
Oph became $\Delta V$$\sim$0.85~mag fainter and $\Delta V-I_{\rm
C}$$\sim$0.6~mag redder than the smooth extrapolation of the light- and
colour-curves of Figure~1.  By day +8 the inflection was over and V2672 Oph
had returned to its smooth decline.

The nature of this inflection, which lasted for $\sim$3.5 days is unknown. 
The only spectroscopic observation secured during this period is an
H$\alpha$ profile which is of no great help, it follows the line-profile
evolution described later in this paper.  Even if occurring around $t_3$,
when dusty novae start to condense grains in their ejecta, the
interpretation of the inflection in terms of small scale dust condensation
in the ejecta of V2672 Oph conflicts with the fact the grains
should have condensed, grown in size and diluted at a speed much faster than
observed in dusty novae, to complete their photometric life cycle in just
3.5 days.

\subsection{Reddening}

The very red colours of V2672 Oph suggest a high reddening. This is
confirmed by the intensity of the NaI D$_1$, D$_2$ interstellar lines and of
the 5780, 5797, 5850, 6270 and 6614 \AA\ diffuse interstellar bands (DIBs)
recorded in our spectra (see later).

The NaI D$_1$, D$_2$ interstellar lines cannot be used because their
very large equivalent width indicates they are largely over-saturated
\citep[cf.][]{MZ97}.  They must be the result of the blend of several
individual components, as it is reasonable to expect given the line-of-sight
passing close to the Galactic centre.  Our spectra lack high enough
resolution to resolve the individual components and thus to check if
they are individually saturated or not.

The best measurable DIB for V2672 Oph is the one at 6614 \AA\ (identified in
Figure \ref{fig:early}).  It appears superimposed against the very
broad and strong H$\alpha$ profile, which provides a good contrast
background.  The other DIBs are instead recorded onto a weak, and therefore
more noisy continuum, which makes the measurement of their equivalent widths
rather uncertain.  The equivalent width of the 6614 \AA\ DIB on the day
+1.31 spectrum is 0.30 $\pm$0.02 \AA.  By comparison with the relation
$E_{B-V}$=5.3$\times$E.W.(\AA) calibrated by \citet{M10} between reddening
and the equivalent width of the 6614~\AA\ DIB, a reddening
$E_{B-V}$$\sim$1.6 can be derived for V2672 Oph.

\citet{BY87} derived a mean intrinsic colour
$(B-V)_\circ$=+0.23 $\pm$0.06 for novae at maximum, and
$(B-V)_\circ$=$-$0.02 $\pm$0.04 for novae at $t_2$. V2672 Oph displayed
$B-V$=+1.77 at $t_2$, to which would correspond $E_{B-V}$$\sim$1.79. The flat
$B$$-$$V$ evolution in Figure~1, suggests that a similar $B-V$=+1.77 could
hold for V2672 Oph also at maximum brightness, with a corresponding
$E_{B-V}$$\sim$1.54. 

The photometric and spectroscopic estimates of the reddening are in fair
agreement, and their average value
\begin{equation}
E_{B-V}=1.6 ~~\pm 0.1
\end{equation}
is adopted in this paper as the reddening affecting V2672 Oph.

\subsection{Distance and peculiar location in the Galaxy}

Most relations between absolute magnitude and the rate of decline take the
form $M_{\rm max}\,=\,\alpha_n\,\log\, t_n \, + \, \beta_n$.  The most
recent calibration of this relation has been provided by \citet{DD00}. 
Their relation for $t_2(V)$ gives $M(V)$=$-$10.40 for V2672 Oph, and
$M(V)$=$-$10.41 for $t_3(V)$.  Adopting the above $E_{B-V}$=1.6 reddening
and a standard $R_V$=3.1 extinction law, the distance to V2672 Oph would be
21 kpc.  The distance remains reasonably large, even if we adopt the
$M(V)$-$t_2$ relation calibrated by \citet{C88} that gives 16 kpc, or the
$M(V)$-$t_3$ relation by \citet{S57} that provides 17 kpc.  Assigning a
higher weight to the more recent calibration by \citet{DD00}, we take the
average of these determinations
\begin{equation} 
d = 19 ~~\pm 2 ~~{\rm kpc} 
\end{equation} 
as the distance to V2672 Oph. At Galactic coordinates $l$=001.021 and
$b$=+02.535, the line-of-sight to V2672 Oph passes close to the Galactic
centre, crosses the whole Bulge and ends at a galacto-centric distance
larger than that of the Sun.  This is probably a record distance and
position among known novae.  The line-of-sight to V2672 Oph does not
cross any of the known Galactic low-extinction windows through the Bulge
listed by \citet{DSB02}, as also confirmed by inspection of SDSS
images showing a scarcely populated stellar field.

Another peculiarity of V2672 Oph and its position is the $z$=0.8 kpc
height above the galactic plane.  \citet{VL98} found that
He/N novae, such as V2672 Oph (see Section 4), belong to the disk
population of the Galaxy and are located at small heights above the Galactic
plane.  While V2672 Oph is undoubtedly away from the Bulge, its height
above the Galactic plane is difficult to reconcile with the $z$$\leq$100-200
pc found by \citet{VL98} for He/N novae.  The proportion of
He/N novae characterised by high $z$ is becoming uncomfortably large in
comparison with the \citet{VL98} scale height $-$ other recent
high-$z$ examples being V2491 Cyg (= Nova Cyg 2008 N.2) at $z$=1.1 kpc
\citep{MSD10}, and V477 Sct (= Nova Sct 2005 N2), located at $z$=0.6
kpc \citep{MSN06}.

 \begin{figure}
 \includegraphics[width=85mm]{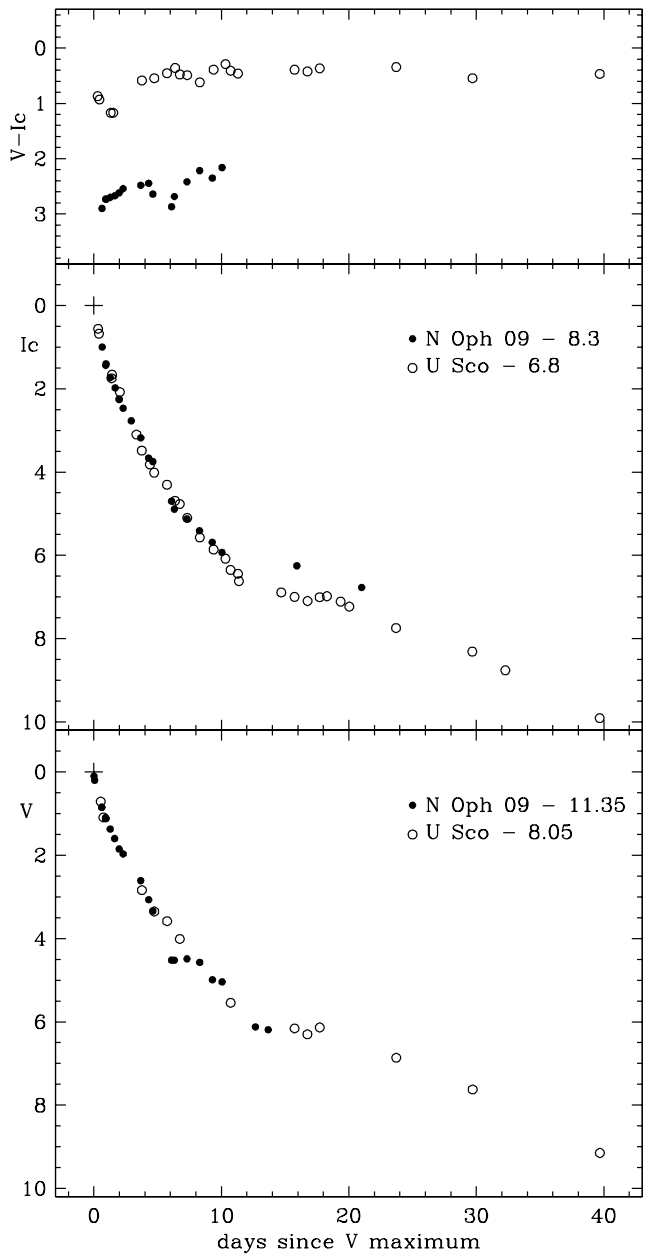}
 \caption{Comparison between the photometric evolutions of V2672 Oph and
 U Sco during its 2010 outburst (from \citet{MDC10}. The $V$ and $I_{\rm C}$
 light-curves are scaled by the given quantities to coincide at their optical
 maxima (see text for details), marked by the cross point at (0,0).}
 \label{fig:fig2}
 \end{figure}

\subsection{The plateau phase and comparison with U Sco}

The light-curves of V2672 Oph and U~Sco (its 2010 outburst) are compared
in Figure~2.  The match is remarkable, with the exception of the
inflection displayed by V2672 Oph around day +5.8 and described in sect. 
3.1 above, which has no counterpart in U Sco.

The strict similarity extends also to later phases, when U~Sco displayed a
plateau phase.  \citet{HKK00,HKK02} and \citet{HKS03} have postulated that a
plateau is characteristic of recurrent novae, and in particular of the U~Sco
subclass of recurrent novae, a position shared by \citet{S10}.  Their idea
is that the plateau is caused by the combination of a slightly irradiated
companion star and a fully irradiated flared disk with a radius 1.4 times
the Roche lobe size.  At the time of the plateau, the outbursting white
dwarf is experiencing stable nuclear burning and supersoft X-ray emission. 
The irradiation of the disk and the companion by a stable central engine
leads to a fairly constant flux added to that steadily declining from the
expanding ejecta.  This leads to a flattening of the light curve (the
plateau), which lasts until the time when the nuclear burning turns off.

The plateau phase of V2672 Oph is well seen in the $I_{\rm C}$
light-curve of Figure~2.  Unfortunately, the observations in the $V$
band stopped due to the faintness of the nova right at the time when
it was entering the plateau phase.  There are just two observations
defining the plateau of V2672 Oph, but they support a close similarity
with the plateau of U~Sco that lasted from day +12 to day +19.  During that
period U~Sco was found to be - as predicted - a bright supersoft X-ray
source \citep{SSP10,OPW10}.

The plateau for V2672 Oph is $\Delta I_{\rm C}$=0.6~mag brighter
than in U~Sco as illustrated by Figure~2, while the $V$-band panel seems to
support instead a similar brightness level.  We do not attach much
significance to the $\Delta I_{\rm C}$ offset.  It could be connected to
differences in the irradiated disks and secondaries between V2672 Oph
and U Sco.

The mass of the white dwarf of U~Sco is believed to be very close to the
Chandrasekhar limit \citep{KH89,HKK00,TDL01}, and thus, if net mass is
being added over time to the white dwarf, it is a prime candidate for
explosion as a Type Ia supernova.  The great similarity of all
observational parameters between U Sco and V2672 Oph leads us to
infer an equally massive white dwarf also in the latter, and an additional
candidate for a future Galactic SNIa.

\subsection{The progenitor}

The progenitor of V2672 Oph had no recorded optical or infrared counterpart
in 2MASS and SDSS surveys.  If the donor star is an M0 giant as in the
recurrent nova RS~Oph, and the parent population of V2672 Oph is the
Galactic disk, then its observed infrared magnitude and colour should be
$K_s$$\sim$12.3 and $J$$-$$K_s$=1.8, or $K_s$$\sim$10.3 and $J$$-$$K_s$=2.0
if the donor star is an M5 giant as in the other recurrent nova T~CrB.

There are only three 2MASS sources within 13 arcsec of the astrometric
position of V2672 Oph.  2MASS 17382110-2644114 lies 5.6 arcsec away and
has $K_s$=12.1 and $J$$-$$K_s$=1.67; 2MASS 17382076-2644079 is 5.5 arcsec
away with $K_s$=10.3 and $J$$-$$K_s$=1.69; 2MASS 17382052-2644184 lies
at a distance of 5.2 arcsec and has $K_s$=13.2 and $J$$-$$K_s$=1.39.  While
the third source appears too blue, the first two could broadly agree with an
M giant at the distance and reddening of V2672 Oph.  Their positions,
however, are not reconcilable with V2672 Oph. \citet{nakano}
lists five independent and accurate measurements for the astrometric
position of the nova.  If $\sigma$ is the r.m.s.  of these five
measurements, then all these three 2MASS sources are more distant than
10$\sigma$ from the mean astrometric position of the nova.

\citet{H85} derived a G0V spectral type for the donor star in U~Sco, shining
in quiescence at $K$=16.45, $J$$-$$K$=0.43.  Closely similar values were
later measured by the 2MASS survey.  Such a donor star for V2672 Oph would
put it far below the detection thresholds of both 2MASS and SDSS surveys.

We may thus conclude that the donor star in V2672 Oph is {\it not} a cool
giant as in RS~Oph and T~CrB, but much more likely a dwarf as in recurrent
novae of the U~Sco type and in most classical novae.  Its orbital period
should therefore be of the order of days, and not months or years.

\section{Spectral evolution}

Only a few spectral lines are usually recognisable on spectra of novae
characterized by very large expansion velocities.  This is caused by the
large blending of individual lines that wash them out into the underlying
continuum energy distribution.  V2672 Oph is no exception to this rule as
illustrated by its spectral evolution presented in Figure~3.  Only H$\alpha$
and OI 8446~\AA\ stand out prominently, while all other lines 
emerge only weakly from the underlying continuum.  Table~3 reports the
absolute fluxes for emission lines that we were able to
recognise and measure with confidence.

The spectra of V2672 Oph classify it among He/N novae, as is usually the
case for fast novae.  The general appearance of the V2672 Oph spectra
is very similar to those of U~Sco, once the reddenings of the two objects
are matched, as illustrated in Figures~3 and 4.

 \begin{figure}
 \includegraphics[width=85mm]{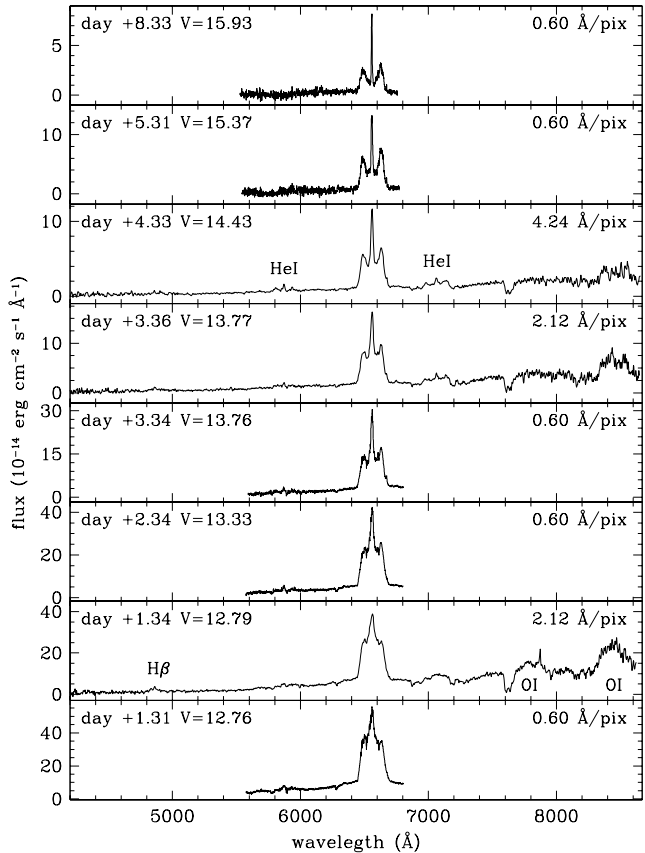}
 \caption{Spectroscopic evolution of V2672 Oph.}
 \label{fig:fig3}
 \end{figure}

 \begin{figure}
 \includegraphics[width=85mm]{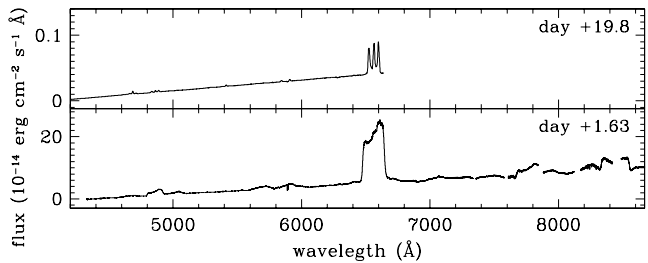}
 \caption{Sample spectra from the 1999 outburst of U Sco to highlight
 the similarities with the spectroscopic evolution of V2672 Oph in
 Figure~3. The U~Sco spectra are from \citet{MZT99}, and their
 reddening has been set equal to $E_{B-V}$=1.6 of V2672 Oph.}
 \label{fig:fig4}
 \end{figure}

The top panel of Figure~1 compares the evolution of the integrated flux of
H$\alpha$ to the photometric one.  The decline rate is identical to that of
the $V$-band flux, while it is slower than for $I_{\rm C}$.  This agrees
with the $V$$-$$I_{\rm C}$ evolution, which is directed toward bluer
$V$$-$$I_{\rm C}$ colours.  The H$\alpha$ contributes less than 10\% to the
$V$-band flux (as proved by integration of the photometric transmission
profile on the spectra as observed and with H$\alpha$ clipped), and this
prevents the equality of $V$-band and H$\alpha$ decline rates from turning
into a circular argument.  The decline of H$\alpha$ flux gradually
accelerated from $F \propto t^{-0.9}$ of early data points in Table~3, to
$F \propto t^{-2.1}$ for the latest ones.  This agrees with the expectation
of $F \propto t^{-3}$ to characterize the latest evolutionary phases when
the H$\alpha$ emissivity settles on the dilution time scale.  The
HeI/H$\alpha$ intensity ratio doubled from day +1.34 to day +4.33, as
expected for an excitation increasing during decline from maximum and
for Case B, still higher than reddening suggests.  The H$\alpha$/H$\beta$
flux ratio remained constant at 35 between these two dates, a very high
value affected by the large reddening to this nova.

The intensity of the OI 8446~\AA\ emission line under normal recombination,
optically thin conditions should be appreciably weaker than the OI 7772
line, 0.6$\times$ its flux.  On day $+$1.34 the reddening-corrected ratio
was 1.63.  The inversion in intensity between the two OI lines is usually
associated with fluorescence pumped by absorption of hydrogen Lyman-$\beta$
photons, as first pointed out by \citet{B47}.  For the Lyman-$\beta$
fluorescence to be effective, the optical depth in H$\alpha$ should be
large, presumably owing to the population of the $n=2$ level by trapped
Lyman-$\alpha$ photons.  The large optical depth in H$\alpha$ is confirmed
also by the $F_{8446}$/$F_{H\alpha}$ flux ratio that under optically thin,
low ionization conditions and typical nova chemical abundances should be
quite low, of the order of $\sim$10$^{-3}$ \citep{SWT77}.  The
reddening corrected ratios for days +1.34, +3.36 and +4.33 are respectively
0.14, 0.18 and 0.11, indicating a persisting large optical depth in
H$\alpha$ during the pre-plateau evolution of V2672 Oph.

A striking feature of V2672 Oph, and once again a matter of close similarity
with U~Sco, is the very large width of its emission lines and how it evolved
with time.  The temporal dependence of the full width at zero intensity
(FWZI) of the H$\alpha$ emission in V2672 Oph is shown in Figure~5, where
the equivalent data for U~Sco are taken from \citet{MZT99}.  The temporal
behaviour for the two novae is identical, with a linear decline of $\sim$250
\kms\ per day, and a starting value of 12,000 \kms\ for V2672 Oph, about
2,000 \kms\ larger than in U Sco and 4,000 \kms\ larger than in RS Oph
\citep{MU07}.  The FWZI depends mainly on the outermost ejecta, those
ejected with the largest velocity.  Compared with the inner ejecta,
the outer ones experience a more rapid decline in electron density and a
large dilution of the ionizing radiation from the central source. 
Consequently the emission of the faster moving ejecta declines faster than
that of the inner ejecta, which results in a progressive reduction in the
width of the emission lines.

 \begin{table}
 \centering
 \caption{Integrated fluxes of emission lines (in units of 10$^{-13}$ erg
          cm$^{-2}$ s$^{-1}$ \AA$^{-1}$).}
 \begin{tabular}{@{~~~}c@{~~~}c@{~~~}c@{~~~}c@{~~~}c@{~~~}c@{~~~}c@{~~~}c@{~~~}c@{~~~}}
             \hline
 day & H$\gamma$ & H$\beta$ & NII & HeI & H$\alpha$ & HeI & OI & OI \\
     &           &          & 5675&5876 &           &7065 & 7774&8446\\
             \hline
 +1.31  &      &      &      &      & 443  &      &      &       \\
 +1.34  & 1.83 & 12.2 & 6.05 & 11.2 & 422  & 21.7 &  80  & 186   \\
 +2.34  &      &      &      &      & 255  &      &      &       \\
 +3.34  &      &      &      &      & 161  &      &      &       \\
 +3.36  &      & 3.75 &      & 5.45 & 140  & 15.0 &      &  80   \\
 +4.33  &      & 2.53 &      & 4.32 & 91   & 13.0 &      &  32   \\
 +5.31  &      &      &      &      & 63   &      &      &       \\
 +8.33  &      &      &      &      & 24   &      &      &       \\
             \hline
 \end{tabular}
 \label{tb:tab3} 
 \end{table} 

 \begin{figure}
 \includegraphics[width=85mm]{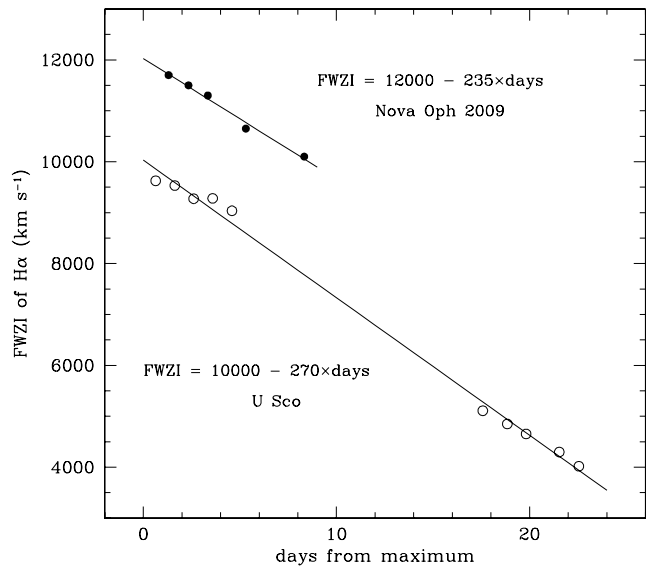}
 \caption{Comparison between the decline with time of the FWZI of H$\alpha$
 for V2672 Oph and U Sco.}
 \label{fig:fig5}
 \end{figure}

 \begin{table*}
 \centering
 \caption{Results of Gaussian fitting to observed spectra. Shown are the
 observation dates, including time since outburst, and individual components
 FWHM and radial velocity displacement in \kms. Components
 N. 1 and 4 relate to the polar blobs, N. 2 to the equatorial ring
 and N. 3 to the prolate structure described in sect 6.2.}
 \begin{tabular}{lcccccccc}
 \hline
 day & \multicolumn{2}{c}{Component 1} & \multicolumn{2}{c}{Component 2} 
     & \multicolumn{2}{c}{Component 3} & \multicolumn{2}{c}{Component 4} \\
     & FWHM & Radial  & FWHM & Radial  & FWHM & Radial  & FWHM & Radial \\
 \hline
 +2.34      & 2128 & -3422 & 982 & -177 & 3791 & -177 & 2258 & 3251 \\
 +3.34      & 2152 & -3466 & 844 & -220 & 3717 & -220 & 2172 & 3116 \\
 +5.31      & 2310 & -3425 & 528 & -225 & 3083 & -225 & 2319 & 3066 \\
 +8.33      & 2283 & -3361 & 333 & -251 & 3598 & -251 & 2314 & 2994 \\
 \hline
 \end{tabular}
 \label{tb:gauss}
 \end{table*}

\section{On the probable nature of V2672 Oph as a recurrent nova}

Such a strict photometric and spectroscopic similarity with U~Sco leads us
to speculate that V2672 Oph is itself a recurrent nova.  However, only one
outburst has been recorded for V2672 Oph, but it is highly probable that
several others have been missed due to the following reasons: (1) V2672 Oph
lies less than 4$^\circ$ away from the ecliptic, and thus suffers from
periods of long seasonal invisibility due to the conjunction with the Sun. 
Furthermore, at every lunation it also suffers from the proximity to
the Moon; (2) its $\delta$ = $-$26$^\circ$44$^\prime$ southern declination
means that it is only observable for a brief fraction of the year by
northern latitudes where most observers have been historically concentrated;
(3) the faint peak brightness attained by V2672 Oph ($V$$\sim$11.35,
$B$$\sim$13.1) means that it would have remained outside the range of
detection by patrol surveys until very recently, when the introduction of
large format CCDs allowed coverage of significant areas of the sky with
short focal length telescopes.  In fact, V2672 Oph was discovered by K. 
Itagaki using a 0.21-m f/3 reflector, a light-collector far more sensitive
that the usual digital SLR cameras used by Japanese amateurs to discover
Galactic novae; and (4) a decline time $t_2(V)$=2.3 days means that
the nova would have returned below the threshold for discovery
($V$$\sim$14) in just three days.

Given all these restrictions, it is indeed surprising that even
this outburst of V2672 Oph has been discovered.  The central region of the
Galaxy has been imaged many times especially by amateurs looking for
impressive pictures.  It is quite possible that other outbursts of V2672 Oph
lie undetected on such archive images, especially those imaging into red
wavelengths.  A devoted search is highly encouraged.  It is less probable
that an outburst could be discovered by inspection of historical plate
archives.  In fact, the $B$$\sim$13.1 mag attained at maximum, places V2672
Oph below the limiting magnitude of most patrol plates
collected worldwide in the past.

\section{Morpho-kinematical modelling of the H$\alpha$ profile}

Using the morpho-kinematical code {\it Shape}\footnote{Available from
http://bufadora.astrosen.unam.mx/shape/} \citep[Version 3.56,][]{SL06,SKW10}
we have analysed and disentangled the three-dimensional geometry and
kinematic structure of the early outburst spectra of V2672 Oph.  {\it Shape}
was originally developed to model the complex structures of planetary
nebulae and is based on computationally efficient mathematical
representations of the visual world which allows for the construction of
objects placed at any orientation in a cubic volume.  {\it Shape} has been
developed so far for modelling optically thin environments, therefore, one
must make the assumption that for what is observed the optical depth is low
and that absorption has not altered greatly the shape of the profile.

The adopted model is based on previous studies of classical novae, which
explored the structures of classical novae from resolved optical imaging and
hydrodynamical modelling \citep[e.g.][]{SOD95,LOB97}.

\subsection{Initial information}

We performed Gaussian fitting using the IRAF task {\tt SPECFIT} on days
$+$2.34 through $+$8.33 after outburst.  These allow us to decompose
different Gaussian components of the H$\alpha$ line and retrieve information
such as the FWHM of likely components and their radial velocity
displacements (Table \ref{tb:gauss}).  We note the presence of the DIB at
6614 \AA (see Figure \ref{fig:early} and sect.  3.2), which is not
considered during detailed modelling given its small equivalent width.

The values derived in Table \ref{tb:gauss} are used to find the displacement
of the system from the rest wavelength of the H$\alpha$ line and to
determine the size of the remnant using the values for the FWHM and radial
velocity displacement of component 1.

The photocentric velocity of the H$\alpha$ profiles in Figures
\ref{fig:aug24} and \ref{fig:early} decreases from $-$19 \kms\ on day +2.34
to $-$146 \kms\ on day +8.33.  Similarly, the velocity of the fitted
equatorial ring and prolate component in Table \ref{tb:gauss} decreases from
$-$177 to $-$251 \kms.  These velocities are probably unrelated to the
systemic velocity of the nova.  The systemic velocity would be
such a small fraction (less than 2\%) of the huge width of H$\alpha$ to make
unrealistic any hope to detect it: velocity offsets are likely to appear for
tiny deviations from perfect point symmetry of the H$\alpha$ emissivity
distribution within the ejecta.  In addition, any absorption within the
ejecta would decrease the flux received from the receding part, introducing
a blue-shift of the photometric barycenter of the line.

\subsection{Modelling}

We used information in Table \ref{tb:gauss} to guide our initial
modeling and e.g. constrain a physical size for the remnant.  The
derivation of the proposed structure of the system was aided from work being
undertaken by the authors for V2491 Cyg (Ribeiro et al.  2010, in
preparation).  Here we explore a combination of structures with polar blobs
and equatorial ring, prolate structure with equatorial rings, prolate
structure with tropical rings \citep[e.g.][]{H72,S83,SOD95} and a structure
similar to RS Ophiuchi as suggested by \citet{RBD09}.  The parameter space
is sampled for inclinations from 0 to 90\degree, and velocities from 100 to
8000 \kms.  Our models of V2491 Cyg produced synthetic spectra for various
combinations of parameters and this allowed us to compare the spectra
observed here to initially place constraints on the overall structure.

 \begin{figure}
 \includegraphics[width=85mm]{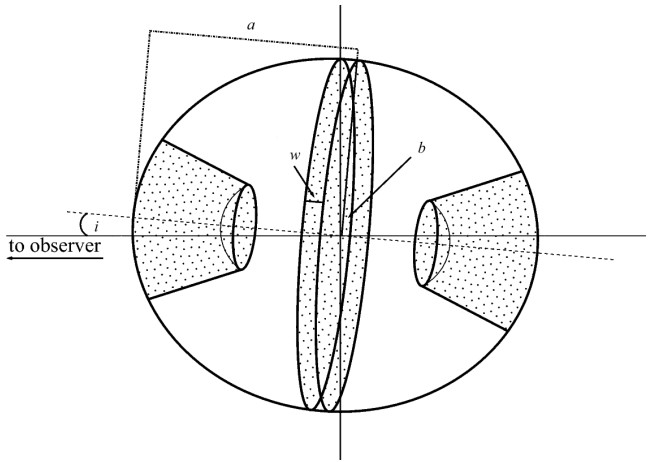}
 \caption{Two-dimensional representation of the three-dimensional expanding
 structure of the V2672 Oph remnant.  The overall structure is that of a
 prolate system with polar blobs and an equatorial ring.  The shaded
 areas represent areas of high density. The semi-major axis ($a$) is larger
 than the semi-minor axis ($b$) by 15\%.  The width of the ring is designated
 $w$.  The inclination of the system ($i$) is defined as the angle between
 the line of sight to the observer and the semi-major axis.}
 \label{fig:model}
 \end{figure}

We note that there is evidence of significant H$\alpha$ optical depth at
early time as derived in Section 4 above.  The modelling of the V2672 Oph
spectra thus initially focused on day $+$8.33 after outburst because at this
time we assume that the H$\alpha$ line is least optically thick. 
In fact, the symmetry of the line is compatible with being optically thin
at this time. The assumed physical size of the object was taken only from
component 1 in Table \ref{tb:gauss} because it is assumed that the blue side
of the line suffers less self-absorption than the red.

Work by \citet{SOD95}, and later updated in \citet{B02}, of several novae
suggested that the remnants could be classified into two broad groups
defined by the speed class of the nova.  \citet{SOD95} 
showed that the axial ratio (the ratio of the lengths of the major and
minor axes of the nova shell) and the speed class ($t$$_3$ time) were
correlated.  It was shown that novae with fast decline from maximum light
demonstrate a lower degree of remnant shaping than slow novae and therefore
we account for this when considering the overall geometry of our system.  A
ratio of 1.15 for the semi-major axes versus semi-minor axis was taken as an
estimate from the nova's speed class.

The suggested structure for the remnant was that of polar blobs and an
equatorial ring with a low density prolate structure surrounding these two. 
The low density prolate structure is used to account for some of the lower
velocities observed in the spectra.  We sample the whole of parameter space
for the inclination of the system from 0 to 90\degree\ (in 1\degree\ steps),
where an inclination of 0\degree\ is for the material expanding along the
semi-major axis being in the line of sight (Figure \ref{fig:model}).  We
also explored the velocity range available for nova explosions from 100 to
8000 \kms\ (in 100 \kms\ steps), assuming a Hubble-flow like velocity
field.

\subsection{Results}

\begin{figure}
\includegraphics[width=84mm]{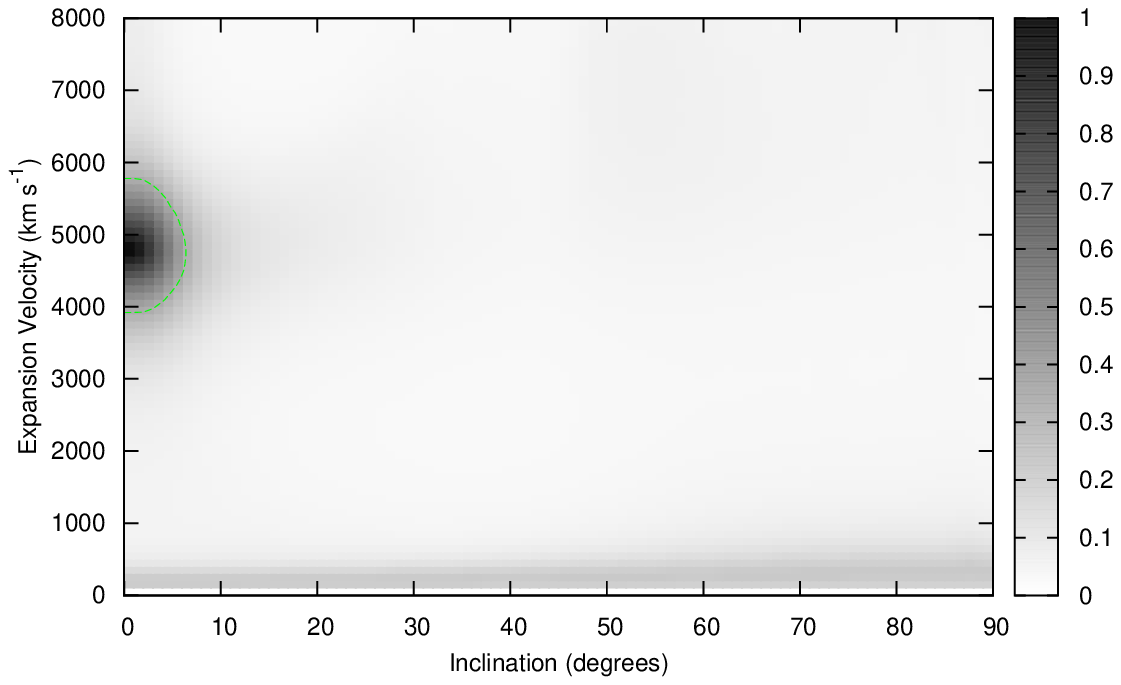}
\includegraphics[angle=270, width=84mm]{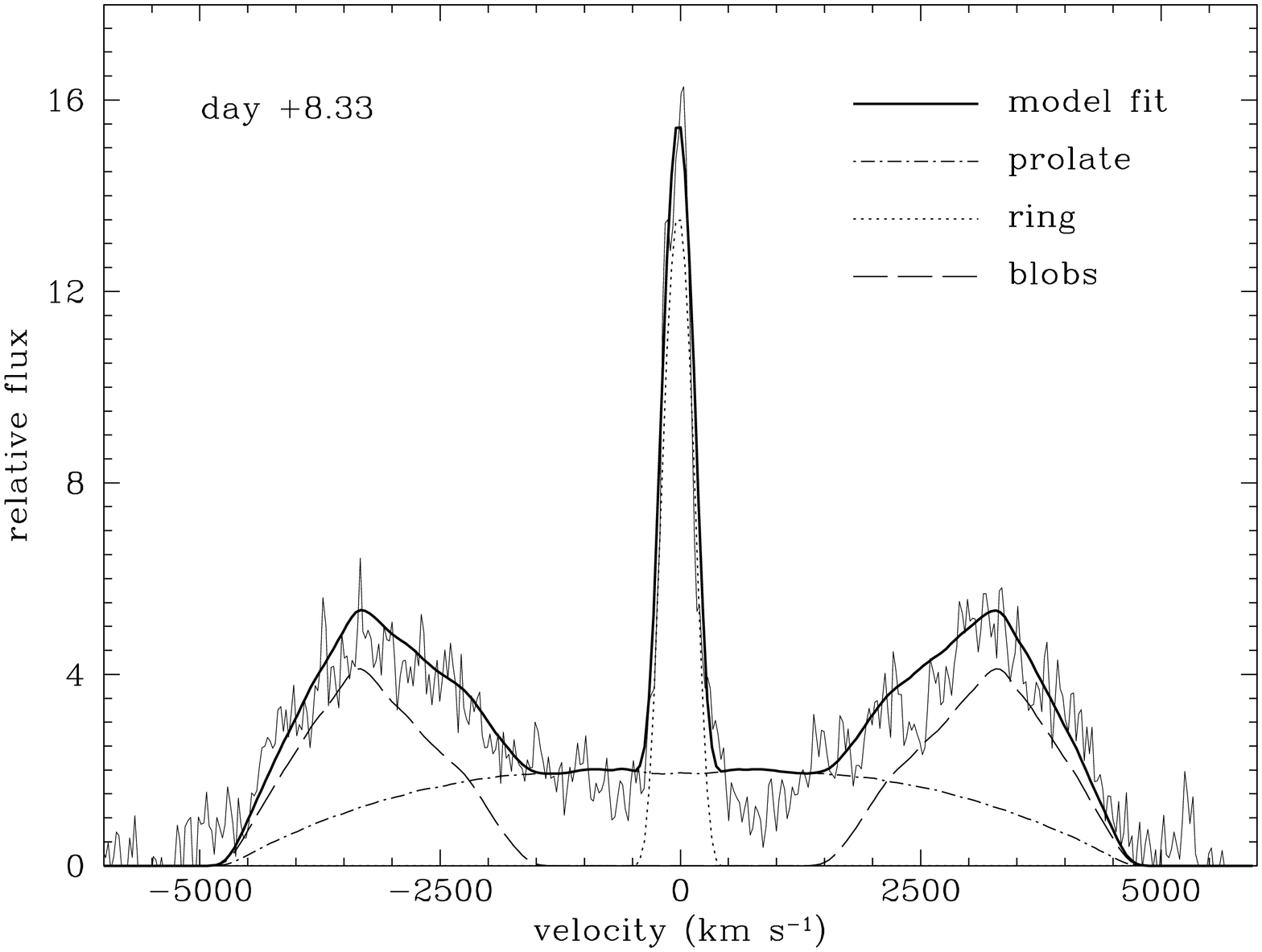}
\caption{Top -- image displaying results of a $\mathcal{X}^2$ fit comparing
the observed spectrum with the model spectrum for different inclinations and
velocities.  The grey scale represents the probability that the observed
$\mathcal{X}^2$ value is correct.  Also shown is the 1$\sigma$ level contour
(dashed green line).  Bottom -- the observed spectrum for day +8.33 and the best
fit model spectrum.  The components of the model fit (prolate region, polar
blobs and equatorial ring) are also shown.}
\label{fig:aug24}
\end{figure}

The observed spectrum and the model spectra for each of the
inclinations and velocities were compared to find the best fit via a
$\mathcal{X}^2$ test (Figure~\ref{fig:aug24}, top).  We derive the best fit
inclination and maximum expansion velocity (at the poles) as
0$\pm$6\degree and 4800$^{+900}_{-800}$ \kms\ respectively (1$\sigma$
confidence intervals).

Figure \ref{fig:aug24} also shows the best fit model spectra and the
contribution of each individual component to the overall model spectrum. 
Just taking a model with polar blobs and an equatorial ring did not match
the overall line profile.  The radial thickness of the polar blobs and
equatorial ring were determined by the FWHM shown in Table \ref{tb:gauss}. 
Even adjusting the densities in each component the model spectrum would not
fully replicate the observed spectrum.  We therefore introduced a filled
prolate structure as an additional component (see Figure \ref{fig:model}). 
This would be associated with material ejected more isotropically than that
in the ring or blobs.  Furthermore, a filled structure can be reconciled
with the assumption that at this stage the post-outburst wind phase and the
ejection of the envelope are still ongoing \citep[e.g.][Ribeiro et al. 
2010, in preparation]{VPD02}.

\begin{figure}
\includegraphics[angle=270, width=84mm]{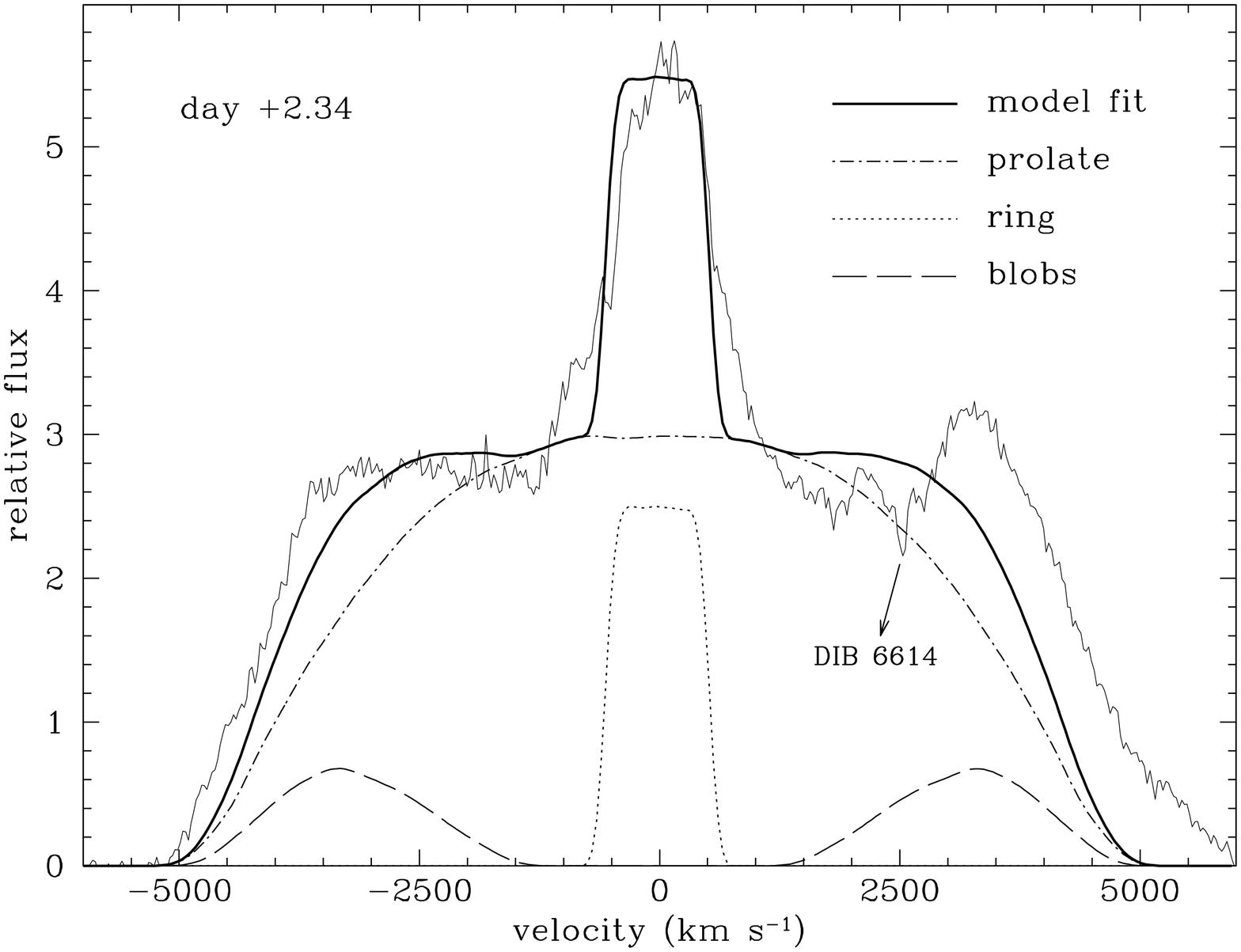}
\includegraphics[angle=270, width=84mm]{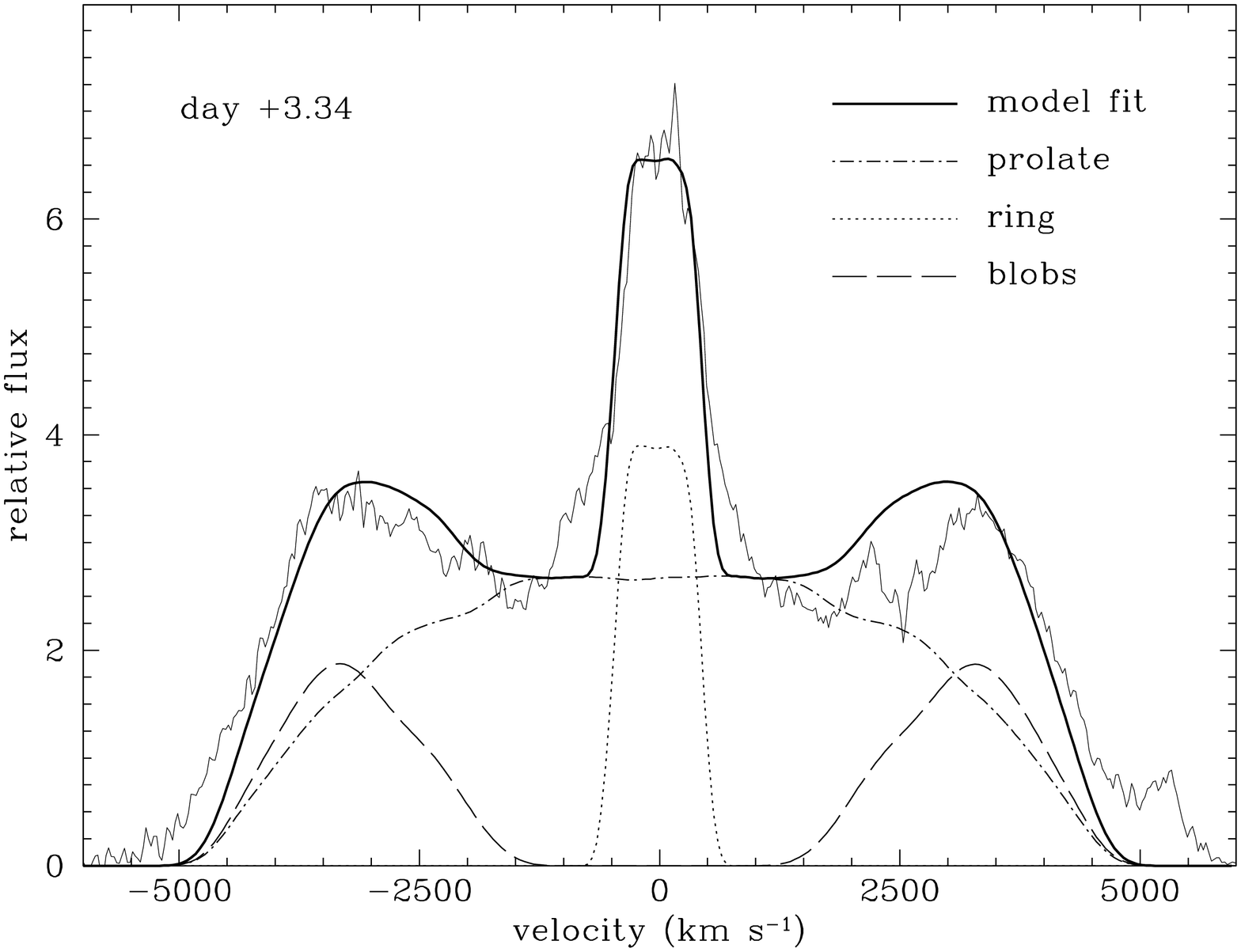}
\includegraphics[angle=270, width=84mm]{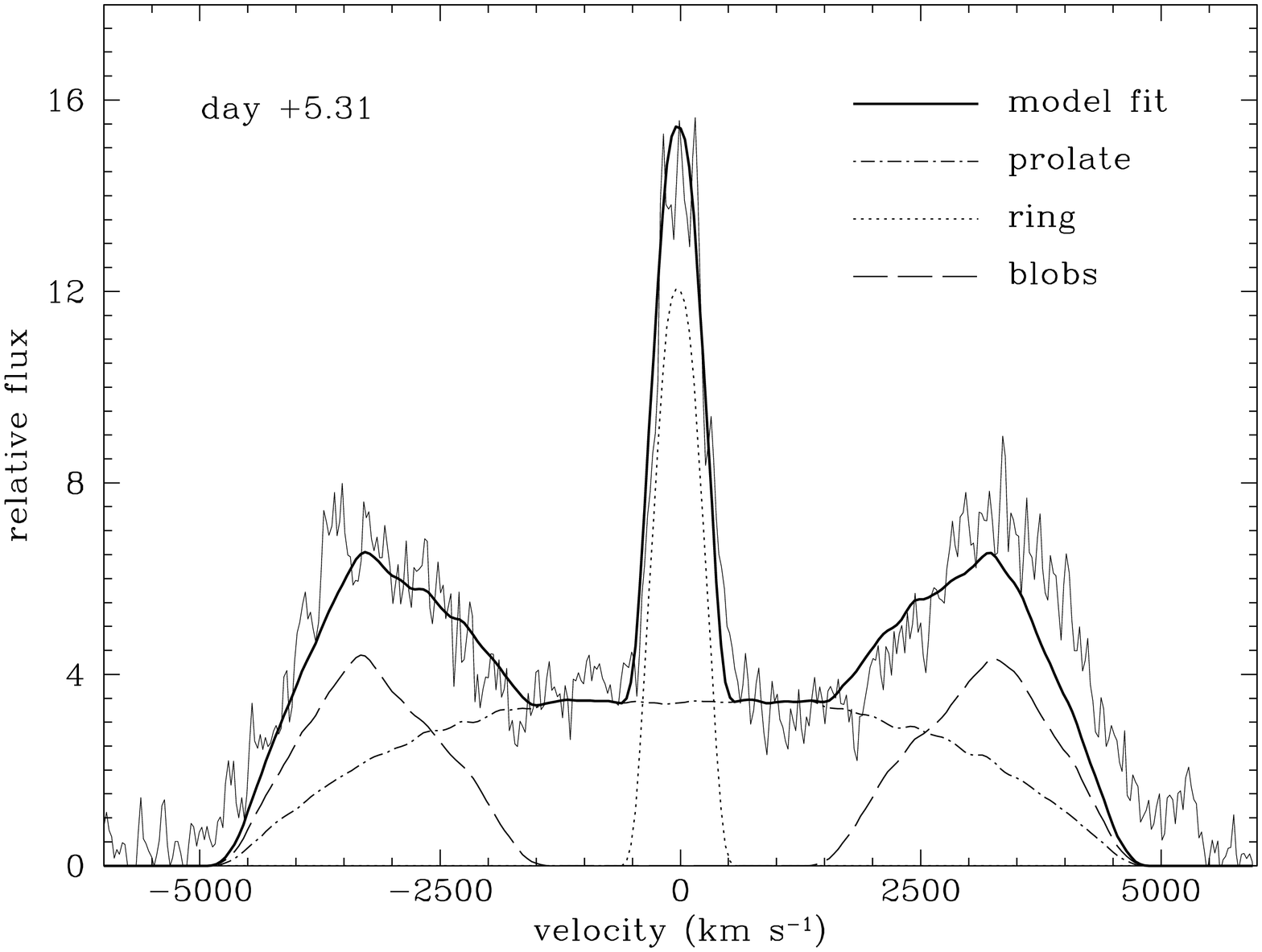}
\caption{Early evolution of the H$\alpha$ profile 
and fitting by model spectra (see Figure \ref{fig:aug24} for further information).}
\label{fig:early}
\end{figure}

We then evolved the overall structure fitted on day $+$8.33 to
earlier times assuming a linear expansion and keeping the inclination and
expansion velocity constant.  As demonstrated in Figure \ref{fig:early}, and
Table \ref{tb:gauss}, the observed spectrum shows the central component has
higher velocities at earlier times, implying that some deceleration has
occurred in this direction as the ejecta expanded.

We then modelled the earlier epoch data (Figure \ref{fig:early}) so that we
keep the polar blobs and prolate structure expanding linearly but the ring
width ($w$) was derived from the values in Table \ref{tb:gauss} for
component 2 (with the ratio of the semi-major to semi-minor axis kept always
as 1.15).  The results are shown in Figure \ref{fig:early}.

The assumption of a decelerating ring replicates the spectra well (Figure
\ref{fig:early}).  The fact that the ring shows a smooth flat top structure
compared to the observed spectra is because we do not include small scale
structure in the model.  There is also some higher velocity material
associated with the wings of the central component that is not reproduced. 
Furthermore, to better replicate the line profiles it was required that we
change the density ratios of the components with time (Table
\ref{tb:density}).  What is evident from the ratio of the densities is that
the prolate structure appears to reduce in density compared with the
other structures while the blobs initially increase in density compared with
the ring then this ratio reduces again at a later time.

\begin{table}
\centering
\caption{Evolution of the implied density ratios for the different model
components.}
\begin{tabular}{lccc}
\hline
day & Blob/Prolate & Blob/Ring & Prolate/Ring \\
\hline
+2.34 & 0.80 & 0.89 & 1.11 \\
+3.34 & 2.36 & 1.60 & 0.69 \\
+5.31 & 4.00 & 1.08 & 0.27 \\
+8.33 & 6.40 & 0.84 & 0.13 \\
\hline
\end{tabular}
\label{tb:density}
\end{table}

\subsection{Discussion}

We have constrained an overall structure, inclination and expansion velocity
for the nebular remnant in V2672 Oph from modelling the H$\alpha$. 
The inclination and maximum expansion velocity have been derived as
0$\pm$6\degree and 4800$^{+900}_{-800}$~\kms \kms\ respectively.  The velocity
derived is consistent with the FWZI at this time (Figure \ref{fig:fig5}). A
structure with polar blobs and an equatorial ring sitting within a prolate
morphology replicates the spectra well.  In classical nova systems however,
a structure with polar blobs and an equatorial ring originates in systems
with low ejecta expansion velocity associated with slow speed class
\citep[e.g.,][]{LOB97}.

The density variations in Table \ref{tb:density} indicate that the prolate
structure quickly reduces in relative density.  While for example, the blobs
initially have lower density than the ring, this ratio then increases,
possibly due to the interaction of the ring material with any pre-existing
equatorial material, only to decrease again later.  Some caution should be
exercised here in over interpretation of these results as more detailed
models, incorporating changes in line optical depths and ionisation
structures are required.

\section{Conclusions and Summary}

We have presented tentative evidence that V2672 Oph is most likely a
recurrent nova similar to U Sco.  This is inferred by their similar
photometric and spectroscopic evolution and by the equally very large
expansion velocities.  Both novae exhibit a very fast decline from maximum,
a plateau phase and a He/N spectrum.  The FWZI of the H$\alpha$
line in V2672 Oph and U Sco declined linearly with a closely similar
slope.  At a distance of 19 kpc and a reddening $E_{B-V}$+1.6, the
line-of-sight to V2672 Oph passes close to the Galactic centre, crosses the
whole Bulge and ends at a galacto-centric distance larger than that of the
Sun.  This is probably a record distance and position among known novae. 
The absence of a 2MASS counterpart suggests the progenitor is not a cool
giant, as in RS Oph and T CrB, but much more likely a dwarf as in U Sco type
of recurrent novae.  Given the southern declination, the faintness at
maximum, the extremely rapid decline and its close proximity to the
Ecliptic, it is quite possible that previous outbursts of V2672 Oph have
been missed.

In terms of the morpho-kinematical modelling (a) if V2672 Oph is a recurrent
nova the shaping mechanism may be different from that of classical novae. 
For example, \citet{RBD09} find RS Oph to have an axial ratio of 3.85 and
compared to its fast decline from maximum, the relationship originally found
by \citet{SOD95} breaks down.  The same could be happening here, except that
V2672 Oph has a main-sequence or subgiant secondary so ejecta/wind
interactions such as in RS Oph may not be important, (b) at this early time
the formation of the remnant may still be taking place and at later times
the remnant could appear more spherical.  We note however that most of the
shaping occurs in the first few hours in classical novae \citep{LOB97}. 
This points to the need for better modelling of the evolution of classical
novae and recurrent novae from the early to late times.  Such work is
however greatly aided if we can ultimately spatially resolve their remnants.

\section*{Acknowledgments}

The authors would like to thank W. Steffen and N. Koning for valuable
discussions on the use of {\it Shape} and adding special features to the
code.  We would like also to thank M.  Darnley for useful discussions
and the ANS Collaboration observers S.  Dallaporta, A.  Frigo, A.  Siviero,
P.  Ochner, S.  Moretti, A.  Maitan, S.  Tomaselli, P.  Valisa and V.  Luppi
for providing part of the data discussed in this paper.  The authors also
would like to thank the anonymous Referee for useful comments and Nick
Almond for assistance in drawing Figure 6.  VARMR is funded by an STFC
studentship.

\label{lastpage}


\begin{thebibliography}{99}

\bibitem[\protect\citeauthoryear{Ayani et al.}{2009}]{AMH09} 
Ayani K., Murakami N., Hata K., Tanaka A., Tachibana M., Kanda A., 2009, 
CBET, 1911, 1 


\bibitem[\protect\citeauthoryear{Bode}{2002}]{B02} Bode 
M.~F., 2002, AIPC, 637, 497 


\bibitem[\protect\citeauthoryear{Bode et al.}{2006}]{BOO06} 
Bode M.~F., et al., 2006, ApJ, 652, 629 


\bibitem[\protect\citeauthoryear{Bowen}{1947}]{B47} Bowen 
I.~S., 1947, PASP, 59, 196 


\bibitem[\protect\citeauthoryear{Cohen}{1988}]{C88} Cohen 
J.~G., 1988, ASPC, 4, 114 


\bibitem[\protect\citeauthoryear{della Valle 
\& Livio}{1998}]{VL98} della Valle M., Livio M., 1998, ApJ, 506, 818 


\bibitem[\protect\citeauthoryear{della Valle et 
al.}{2002}]{VPD02} della Valle M., Pasquini L., Daou D., Williams R.~E., 2002, A\&A, 390, 155 


\bibitem[\protect\citeauthoryear{Downes 
\& Duerbeck}{2000}]{DD00} Downes R.~A., Duerbeck H.~W., 2000, AJ, 120, 2007 


\bibitem[\protect\citeauthoryear{Dutra, Santiago, 
\& Bica}{2002}]{DSB02} Dutra C.~M., Santiago B.~X., Bica E., 2002, A\&A, 381, 219 


\bibitem[\protect\citeauthoryear{Eyres et al.}{2009}]{EOB09} 
Eyres S.~P.~S., et al., 2009, MNRAS, 395, 1533 


\bibitem[\protect\citeauthoryear{Hachisu et 
al.}{2000}]{HKK00} Hachisu I., Kato M., Kato T., Matsumoto 
K., 2000, ApJ, 528, L97 


\bibitem[\protect\citeauthoryear{Hachisu et 
al.}{2002}]{HKK02} Hachisu I., Kato M., Kato T., Matsumoto 
K., 2002, ASPC, 261, 629 


\bibitem[\protect\citeauthoryear{Hachisu, Kato, 
\& Schaefer}{2003}]{HKS03} Hachisu I., Kato M., Schaefer B.~E., 2003, ApJ, 584, 1008 


\bibitem[\protect\citeauthoryear{Hanes}{1985}]{H85} Hanes 
D.~A., 1985, MNRAS, 213, 443 


\bibitem[\protect\citeauthoryear{Hutchings}{1972}]{H72} 
Hutchings J.~B., 1972, MNRAS, 158, 177 


\bibitem[\protect\citeauthoryear{Kato 
\& Hachisu}{1989}]{KH89} Kato M., Hachisu I., 1989, ApJ, 346, 424 


\bibitem[\protect\citeauthoryear{Krauss Hartman, Rupen, 
\& Mioduszewski}{2009}]{HRM09} Krauss Hartman M.~I., Rupen M.~P., Mioduszewski A.~J., 2009, ATel, 2195, 1 


\bibitem[\protect\citeauthoryear{Landolt}{1983}]{L83} 
Landolt A.~U., 1983, AJ, 88, 439 


\bibitem[\protect\citeauthoryear{Landolt}{1992}]{L92} 
Landolt A.~U., 1992, AJ, 104, 340 


\bibitem[\protect\citeauthoryear{Landolt}{2009}]{L09} 
Landolt A.~U., 2009, AJ, 137, 4186 


\bibitem[\protect\citeauthoryear{Lloyd, O'Brien, 
\& Bode}{1997}]{LOB97} Lloyd H.~M., O'Brien T.~J., Bode M.~F., 1997, MNRAS, 284, 137 


\bibitem[\protect\citeauthoryear{Munari 
\& Zwitter}{1997}]{MZ97} Munari U., Zwitter T., 1997, A\&A, 318, 269 


\bibitem[\protect\citeauthoryear{Munari et 
al.}{1999}]{MZT99} Munari U., et al., 1999, A\&A, 347, L39 


\bibitem[\protect\citeauthoryear{Munari et 
al.}{2006}]{MSN06} Munari U., Siviero A., Navasardyan H., Dallaporta S., 2006, A\&A, 452, 567 

\bibitem[\protect\citeauthoryear{Munari et al.}{2007}]{MU07} 
Munari U., et al., 2007, BaltA, 16, 46 

\bibitem[\protect\citeauthoryear{Munari et al.}{2009}]{MSO09} 
Munari U., Saguner T., Ochner P., Siviero A., Maitan A., Valisa P., 
Dallaporta S., Moretti S., 2009, CBET, 1912


\bibitem[\protect\citeauthoryear{Munari, Dallaporta, 
\& Castellani}{2010}]{MDC10} Munari U., Dallaporta S., Castellani F., 2010, IBVS, 5930


\bibitem[\protect\citeauthoryear{Munari et al.}{2010}]{MSD10} Munari U., Siviero A., Dallaporta S., Cherini G., Valisa P., Tomassella L., 2010,
New Astronomy, to be submitted


\bibitem[\protect\citeauthoryear{Munari}{2010}]{M10} Munari U., 2010, PASP, to be submitted


\bibitem[\protect\citeauthoryear{Nakano, Yamaoka, 
\& Kadota}{2009}]{nakano} Nakano S., Yamaoka H., Kadota K., 2009, CBET, 1910, 1 


\bibitem[\protect\citeauthoryear{O'Brien, Lloyd, 
\& Bode}{1994}]{OLB94} O'Brien T.~J., Lloyd H.~M., Bode M.~F., 1994, MNRAS, 271, 155 


\bibitem[\protect\citeauthoryear{O'Brien et 
al.}{2006}]{OBP06} O'Brien T.~J., et al., 2006, Nature, 442, 
279 


\bibitem[\protect\citeauthoryear{Osborne et 
al.}{2010}]{OPW10} Osborne J.~P., et al., 2010, ATel, 2442, 1 


\bibitem[\protect\citeauthoryear{Padin, Davis, 
\& Bode}{1985}]{PDB85} Padin S., Davis R.~J., Bode M.~F., 1985, Nature, 315, 306 


\bibitem[\protect\citeauthoryear{Ribeiro et 
al.}{2009}]{RBD09} Ribeiro V.~A.~R.~M., et al., 2009, ApJ, 
703, 1955 


\bibitem[\protect\citeauthoryear{Schaefer}{2010}]{S10} 
Schaefer B.~E., 2010, ApJS, 187, 275 


\bibitem[\protect\citeauthoryear{Schlegel et 
al.}{2010}]{SSP10} Schlegel E.~M., et al., 2010, ATel, 2430, 
1 


\bibitem[\protect\citeauthoryear{Schmidt}{1957}]{S57} 
Schmidt T., 1957, ZA, 41, 182 


\bibitem[\protect\citeauthoryear{Schwarz et 
al.}{2009}]{SOP09} Schwarz G.~J., et al., 2009, ATel, 2173, 1 


\bibitem[\protect\citeauthoryear{Slavin, O'Brien, 
\& Dunlop}{1995}]{SOD95} Slavin A.~J., O'Brien T.~J., Dunlop J.~S., 1995, MNRAS, 276, 353 


\bibitem[\protect\citeauthoryear{Solf}{1983}]{S83} Solf J., 
1983, ApJ, 273, 647 


\bibitem[\protect\citeauthoryear{Steffen 
\& L{\'o}pez}{2006}]{SL06} Steffen W., L{\'o}pez J.~A., 2006, RMxAC, 26, 30 


\bibitem[\protect\citeauthoryear{Steffen et 
al.}{2010}]{SKW10} Steffen W., Koning N., Wenger S., Morisset 
C., Magnor M., 2010, arXiv, arXiv:1003.2012 


\bibitem[\protect\citeauthoryear{Strittmatter et 
al.}{1977}]{SWT77} Strittmatter P.~A., et al., 1977, ApJ, 
216, 23 


\bibitem[\protect\citeauthoryear{Thoroughgood et 
al.}{2001}]{TDL01} Thoroughgood T.~D., Dhillon V.~S., 
Littlefair S.~P., Marsh T.~R., Smith D.~A., 2001, MNRAS, 327, 1323 


\bibitem[\protect\citeauthoryear{van den Bergh 
\& Younger}{1987}]{BY87} van den Bergh S., Younger P.~F., 1987, A\&AS, 70, 125 

\end{thebibliography}
\end{document}